\documentclass[conference]{IEEEtran}
\IEEEoverridecommandlockouts

\usepackage{cite}
\usepackage{amsmath,amssymb,amsfonts}
\usepackage{algorithmic}
\usepackage[hidelinks]{hyperref}
\usepackage{graphicx}
\usepackage{subfig}
\usepackage{textcomp}
\usepackage{xcolor}
\usepackage{caption}
\usepackage[numbers]{natbib}
\def\BibTeX{{\rm B\kern-.05em{\sc i\kern-.025em b}\kern-.08em
		T\kern-.1667em\lower.7ex\hbox{E}\kern-.125emX}}
\begin{document}
	
	\title{CEEMS: A Resource Manager Agnostic Energy and Emissions Monitoring Stack}
	
	\author{\IEEEauthorblockN{1\textsuperscript{st} Mahendra Paipuri}
		\IEEEauthorblockA{\textit{Institut du Développement et des Ressources en Informatique Scientifique (IDRIS)} \\
			\textit{National Centre for Scientific Research (CNRS)}\\
			Orsay, France \\
			mahendra.paipuri@cnrs.fr}
	}
	
	\maketitle
	
	\begin{abstract}
		With the rapid acceleration of ML/AI research in the last couple of years, the energy consumption of the Information and Communication Technology (ICT) domain has rapidly increased. As a major part of this energy consumption is due to users' workloads, it is evident that users need to be aware of the energy footprint of their applications. Compute Energy \& Emissions Monitoring Stack (CEEMS) has been designed to address this issue. CEEMS can report energy consumption and equivalent emissions of user workloads in real time for HPC and cloud platforms alike. Besides CPU energy usage, it supports reporting energy usage of workloads on NVIDIA and AMD GPU accelerators. CEEMS has been built around the prominent open-source tools in the observability eco-system like Prometheus and Grafana. CEEMS has been designed to be extensible and it allows the Data Center (DC) operators to easily define the energy estimation rules of user workloads based on the underlying hardware. This paper explains the architectural overview of CEEMS, data sources that are used to measure energy usage and estimate equivalent emissions and potential use cases of CEEMS from operator and user perspectives. Finally, the paper will conclude by describing how CEEMS deployment on the Jean-Zay supercomputing platform is capable of monitoring more than 1400 nodes that have a daily job churn rate of around 20k jobs.
	\end{abstract}
	
	\begin{IEEEkeywords}
		Energy measurement, emissions, monitoring, HPC, cloud, GPU.
	\end{IEEEkeywords}

	\section{Introduction}
	\label{sec:1}
	
	With the advent of climate change, there is an immediate need to reduce and optimize the energy usage of ICT technologies. This is only possible when end users and Data Center (DC) operators are made aware of their energy and environmental footprint. Various options are available to report energy usage of compute workloads on HPC and cloud platforms. The most accurate and less intrusive method is to use external watt meters. However, this requires installing additional infrastructure which might not be a scalable option. Moreover, external watt meters measure the consumption of the entire compute node and the energy consumption data of different processes/components are not readily available. Software-based power meters have gained a lot of traction in the last decade and they have different levels of measurement granularity. The downside to software-based power meters is that they are specific to each processor and require detailed knowledge of underlying hardware.
	
	
	Various monitoring platforms have been introduced in the past like Ganglia~\cite{Massie:2004}, XDMoD~\cite{Palmer:2015}, TACC Stats~\cite{Evans:2014} and PIKA~\cite{Dietrich:2020}. These solutions lack either better visualization tools, GPU support, energy usage metrics at the compute workload level or customizable energy estimation framework. Prometheus has become a \emph{de-facto} solution for storing time series data in a Time Series Database (TSDB) in the cloud landscape. The raw data from different data sources can be ingested into Prometheus TSDB and using recording rules~\cite{promdocs:recordingrules}, it is possible to estimate the same derived metric using different rules according to the needs and underlying hardware of the DC. In the current context, this means it is possible to ingest energy usage data from different sources like in-band, out-band measurements, watt meters, \emph{etc.,} and operators can use one or more data sources to estimate the energy usage of each compute workload. Grafana proved to be a versatile solution in the visualization landscape that has strong support for Prometheus. Grafana allows operators to further customize the visualization tools using its plugin eco-system~\cite{grafana:plugintools} to add custom third-party data sources and/or panels.
	
	In the present work, CEEMS~\cite{ceems:2024} which is based on Prometheus and Grafana will be introduced. CEEMS has been designed to be platform agnostic meaning that it can be deployed on an HPC, Openstack or Kubernetes platform that is running on bare-metal to gather energy and emissions data of individual workloads. The rest of the paper is organized as follows: Section~\ref{sec:2} discusses the overall architecture of CEEMS, Section~\ref{sec:3} briefs configuration aspects of CEEMS using its deployment on Jean-Zay supercomputer as an example and finally, Section~\ref{sec:4} concludes the paper with few pointers to future work.
	
	\section{Overview}
	\label{sec:2}
	
	\subsection{Philosophy}
	
	This section explains how different metrics are gathered from different sources in CEEMS.
	
	\paragraph{CPU, memory and IO metrics}
	
	Most of the resource managers like SLURM, Libvirt (for Openstack), and Kubelet (for Kubernetes) use Linux Control Groups (cgroups)~\cite{kernel:cgroups} to manage CPU, memory and IO resources for compute workloads. cgroups have different controllers like CPU, memory, IO, CPUset, PID, \emph{etc.,} which control the usage of that resource. Essentially resource managers create a cgroup for each compute workload (a batch job for the HPC platform, a VM for Openstack, and a pod for Kubernetes) with the resource constraints imposed based on the workload specification. Consequently, cgroups will keep track of accounting information of each workload to ensure the resource usage of a given workload at a given time does not exceed the allocated resources. This accounting information is stored in the \verb|/sys/fs/cgroup| pseudo-filesystem on each compute node. Thus, by walking through the cgroups file system on each compute node, it is possible to gather the accounting information of each compute unit. Types of metrics available by this method depend on the different cgroup controllers that the resource manager supports. The advantage of this approach is gathering workload metrics is very cheap and non-intrusive. Besides cgroups metrics, node-level metrics like total CPU usage, total memory usage, \emph{etc.,} are also collected from \verb|/sys| and \verb|/proc| pseudo-filesystems.
	
	\paragraph{Energy metrics}
	
	The most common and widely used metric for measuring energy usage is the Running Average Power Limit (RAPL)~\cite{intel:sdmvol4} counters introduced by Intel in Sandy Bridge architecture. RAPL reports energy consumption on different levels or power domains.
	Server class hardware comes with intra-node devices which include Baseboard Management Controllers (BMC) that can measure the power consumption of the node. These measurements are exposed to the Operating System (OS) using the IPMI-DCMI~\cite{intel:ipmidcmi} interface. The advantage of energy measurements reported by IPMI-DCMI is that they report the total power consumption of the entire node, unlike RAPL counters which include only certain components of the node. On the other hand, the IPMI-DCMI command is not suitable to use at a high frequency (even for every few seconds) whereas RAPL counters are available at microsecond granularity.
	
	Both RAPL counters and IPMI-DCMI readings are collected in CEEMS. This gives the operators the freedom to choose which metrics to present based on their underlying hardware. It is also possible to mix both metrics which will be discussed in Section~\ref{sec:3}. Both RAPL and IPMI-DCMI report energy usage at the node level and it is necessary to distribute the total energy consumption among different compute workloads running on the node at a given time. It is very difficult to have an exact energy usage of each process as modern processors are very complex with deep pipelines. Kepler~\cite{kepler:powermodel} uses a modeling approach by using current energy usage metrics with pre-trained power models to estimate the energy consumption of individual processes. However, CEEMS uses a simple model by distributing the energy consumption based on the relative CPU time of each workload to the total CPU time of the node, which stays a very good approximation. Scaphandre~\cite{scaphandre:2024} uses a similar approach with RAPL counters to estimate the energy consumption of each process.
	
	\paragraph{Emissions metrics}
	
	Normally equivalent emissions are estimated based on the energy consumption and the emission factor which is defined as equivalent grams of CO$_{2}$ emitted per Kilo Watt Hour energy consumed. Emission factor depends on the current energy mix, \emph{i.e.,} proportion of different sources of energy in the current electricity generation. Energy mix data is dynamic in time and so as are emission factors. OWID provides static emission factor data~\cite{owid:emissions} for different countries estimated based on historical data. Fortunately, more and more countries are providing real-time emission factors like RTE, the French national electricity distributor does for France~\cite{rte:emissions}. Electricity Maps~\cite{emaps:emissions}, a Copenhagen-based company provides real-time emission factors for a lot of countries and offers free API access for non-commercial usage. Currently, CEEMS gathers static emission factor data from OWID and real-time data from RTE and Electricity Maps.
	
	\paragraph{GPU metrics}
	
	Currently, CEEMS supports only NVIDIA and AMD GPU accelerators by leveraging the metrics exposed through NVIDIA DCGM and AMD ROCm interfaces, respectively. For most resource managers including SLURM, the indices of GPU devices that are bound to a workload will not be available post-mortem of the workload. Thus, CEEMS collects and stores the map information of workload ID to GPU indices.
	
	\subsection{Components}
	
	CEEMS loosely follow microservices architecture where three different components work in tandem to monitor the compute workloads. These components are described below:
	
	\paragraph{CEEMS exporter}
	
	This is a Prometheus exporter that runs on each compute node of the cluster. Essentially it is an HTTP server that runs on the node and sends the metrics response to every request in a format understandable by Prometheus. It is very lightweight consuming minimum resources. On average the exporter consumes 15-20 MB of memory and each scrape request takes less than 1 microsecond of CPU time. The exporter contains different collectors which can be enabled or disabled based on needs using CLI options. The exporter supports basic auth and TLS to protect it from DoS/DDoS attacks from malicious users. When using GPU clusters, either DCGM exporter~\cite{dcgm:exporter} or AMD SMI exporter~\cite{amd:exporter} must be deployed alongside the CEEMS exporter to collect GPU metrics.
	
	\paragraph{CEEMS API server}
	
	As discussed earlier, CEEMS uses Prometheus to store time series metrics of user workloads. For large clusters, the collected data points can quickly become exorbitant with time. Although Prometheus is a highly performant TSDB, it is not suitable to make queries that span a long duration. An example of such a query can be the total energy usage of a given user or a project on a given cluster for all the workloads during the last year. CEEMS API server has been created to address this issue by storing the aggregate metrics of compute units, users and projects. CEEMS API server fetches information from two different sources: the underlying resource manager to get a list of compute workloads and Prometheus to fetch the metrics of these compute workloads. CEEMS API server uses a relational DB based on SQLite to store the compute workloads and their aggregated metrics. CEEMS API server also serves as an abstraction layer for different resource managers by defining a unified DB schema to store compute units of different resource managers. CEEMS API server's API documentation~\cite{ceems:apidocs} provides more details on available API endpoints.
	
	\paragraph{CEEMS load balancer}
	
	Although Prometheus and Grafana proved to be an impeccable pair in monitoring and observability, they lack one crucial element: Access Control. There is nothing preventing a malicious user that has read access on a data source in Grafana to query for the metrics of \emph{any} compute workload. This is an undesirable consequence from both security and privacy perspectives. CEEMS Load Balancer (LB) has been created to address this issue. It has the dual functionality of being a reverse proxy and load balancer to Prometheus. CEEMS LB intercepts the query request to the backend Prometheus instance, retrieves the workload unique identifier and checks with the CEEMS API server if the workload belongs to the user who initiated the request. The current user is identified using \verb|X-Grafana-User| header that Grafana sends with every query to the backend data source~\cite{grafana:docs}. If the check is successful, CEEMS LB will pass the request to the backend Prometheus and serve the response. As a load balancer, it can perform load balancing among multiple Prometheus instances using classic strategies like round-robin and least connection. Thus, CEEMS LB provides the missing access control element through its reverse proxy role.
	
	\subsection{Architecture}
	\label{sec:31}
	
\begin{figure*}
		\centerline{\includegraphics[width=\textwidth,trim={0 4cm 0 4cm},clip]{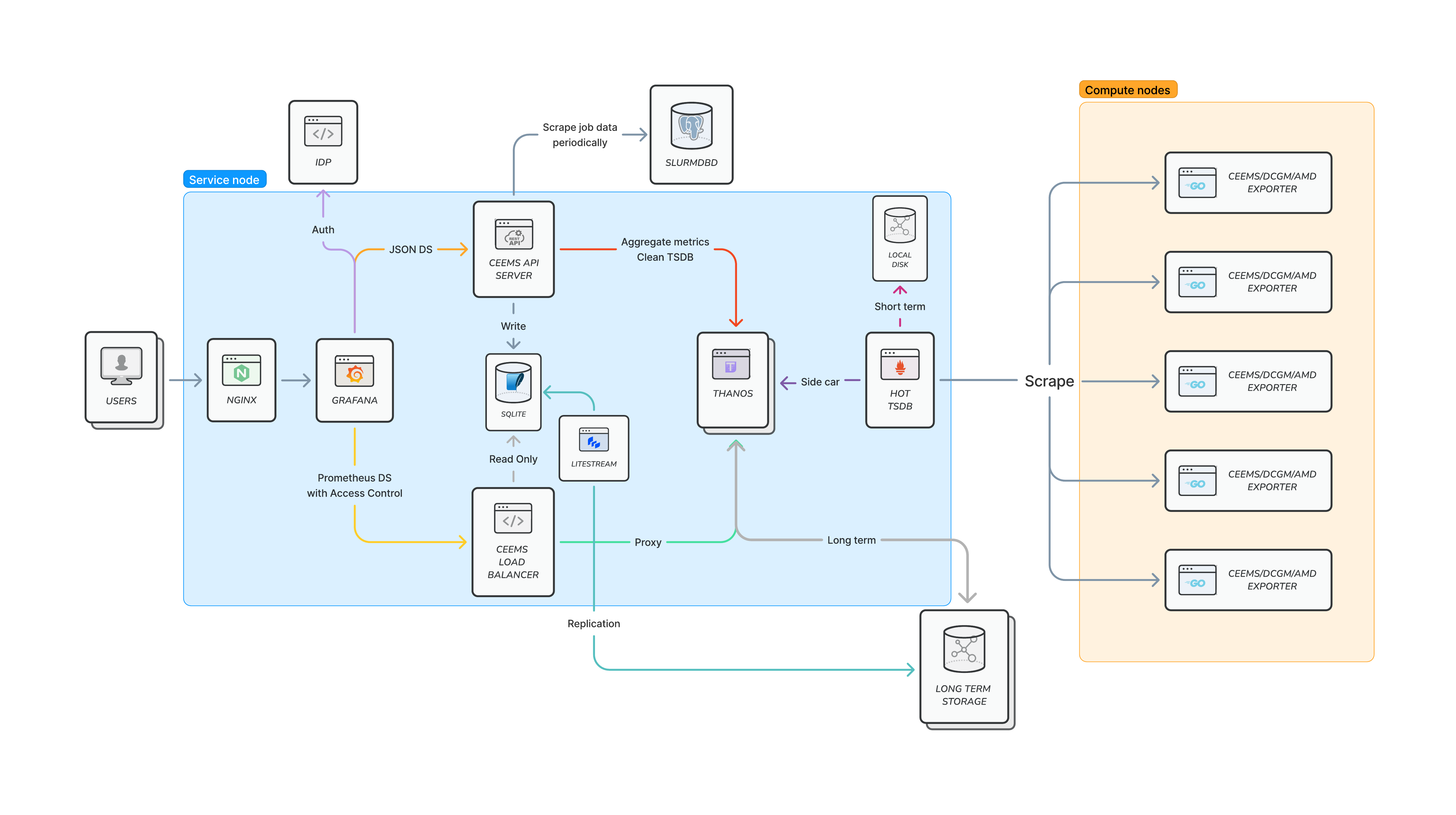}}
		\caption{CEEMS architecture using Thanos and Litestream.}
		\label{fig:ceemsarch}
	\end{figure*}

	Fig.~\ref{fig:ceemsarch} shows a reference architecture of CEEMS in the context of an HPC platform with SLURM as a batch scheduler. 
	Compute nodes of the cluster are represented on the right-hand side where CEEMS and DCGM/AMD exporters are running. A hot TSDB instance will scrape these compute nodes at a configured interval and ingest the metrics of compute workloads on a local disk. Prometheus does not advise to use network file system as a storage backend due to broken POSIX locks implementations on most of the network file systems. This hot TSDB will replicate the data to Thanos~\cite{thanos:intro}, which provides long-term storage capabilities to Prometheus. 
	
	As discussed above, the CEEMS API server fetches the job data from SLURM DBD periodically and populates its own DB based on SQLite. At the same time, the CEEMS API server estimates the aggregate metrics by querying Thanos. It is possible to configure the CEEMS API server to clean up TSDB by removing metrics of workloads that did not last more than the configured cutoff time. This helps in reducing the cardinality~\cite{grafana:cardinalityblog} of metrics. Optionally, SQLite DB can be backed up continuously onto long-term storage using Litestream~\cite{litestream:intro}. CEEMS API server also supports an in-built punctual backup solution at a configured interval.  
	
	CEEMS LB sits in front of Thanos to introspect the query before deciding whether to proxy the query or not. It does so by directly querying the CEEMS API server's DB, when available. If the DB file is not accessible, CEEMS LB makes an API request to the CEEMS API server to check the user's ownership status to the queried workload. 
	
	Finally, Grafana dashboards can be built using Prometheus and CEEMS API servers as data sources. Prometheus data source provides the time series metrics of each compute workload whereas the CEEMS API server data source serves the aggregated metrics and list of compute workloads.
	
	\subsection{Technical details}
	
	The entire CEEMS stack has been implemented in Golang to take advantage of the concurrency model of the programming language. The ease of cross-compiling in Golang is another motivation to choose the language where pre-compiled binaries can be used directly on the target platforms. CEEMS supports different installation methods including containerized deployments, packaged OS and Ansible roles. More details can be found in the installation section of docs~\cite{ceems:docs}. SQLite has been chosen as Relational Data Base Management System (RDBMS) for its simplicity and lack of external dependencies. Moreover, the CEEMS API server do not need concurrent write access as there is only one go routine that writes to DB at a configured interval. This justifies the use of SQLite in the current context. All the CEEMS components can be configured in a single YAML file where each component will read its relevant configuration. All CEEMS components support basic auth and TLS.
	
	
	\section{Deployment on Jean-Zay}
	\label{sec:3}
	
	CEEMS has been deployed on the Jean-Zay supercomputer, which is a heterogeneous system with approximately 1400 compute nodes (Intel and AMD). There are more than 3500 NVIDIA GPUs (V100, A100 and H100) distributed among different partitions of the system. Most of the compute nodes are diskless and contain a mix of Infiniband and Omni-Path networking. 
	
	\subsection{Job energy usage estimation}
	
	As discussed in Section~\ref{sec:1}, an important feature of CEEMS is configurability and customization of energy estimation of compute workloads. CEEMS exporter exports the energy consumption metrics reported by both IPMI-DCMI and RAPL on all compute nodes of Jean-Zay. On Intel compute nodes RAPL reports energy usage counters for both CPU and DRAM whereas on AMD compute nodes, only CPU energy counters are reported by RAPL. There are at least two different types of GPU servers. In one server type, IPMI-DCMI reports the total power usage of the compute node including the power usage by GPUs where whereas the other type does not include GPU power usage in the IPMI-DCMI readings. Each of these cases can be handled elegantly using Prometheus by grouping them in different scrape target groups and defining the recording rules accordingly. 
	
	For instance, when both RAPL CPU and DRAM energy counters are available, the power usage of each job is estimated using the following approach. As most of the compute nodes of Jean-Zay are diskless, the power consumption of local storage is assumed to be zero. Furthermore, 10\% of the total energy consumption is attributed to network~\cite{Dayarathna:2016} and the rest to CPU and DRAM. The energy consumption of other components like memory controllers, registers and PCIe devices are included in the consumption of CPU and DRAM. Power usage of a given job is estimated as in~\eqref{eq:1}
	\begin{multline}
		\label{eq:1}
		P_{job,t} = 0.9 * P_{ipmi,t} * \left(\dfrac{P_{rapl,cpu,t}}{P_{rapl,cpu,t} + P_{rapl,dram,t}}\right) * \left(\dfrac{T_{job,t}}{T_{node,t}}\right) \\
		+ 0.9 * P_{ipmi,t} * \left(\dfrac{P_{rapl,dram,t}}{P_{rapl,cpu,t} + P_{rapl,dram,t}}\right) * \left(\dfrac{M_{job,t}}{M_{node,t}}\right) \\
		+ 0.1 * P_{ipmi,t} * \left(\frac{1}{N_{jobs,t}}\right)
	\end{multline}
	where $P_{job,t}$ is power usage of job, $P_{ipmi,t}$ is power reported by IPMI for entire node, $P_{rapl,cpu,t}$ and $P_{rapl,dram,t}$ power readings from RAPL counters for CPU and DRAM, respectively, $T_{job,t}$ is CPU time of job, $T_{node,t}$ is total CPU time of node, $M_{job,t}$ is memory usage of job, $M_{node,t}$ is memory usage of entire node and $N_{job,t}$ number of jobs running on the node. All the variables are defined at a given time $t$. 
	
	The first two terms in~\eqref{eq:1} represent the CPU and DRAM power usage of the job, respectively. The total power reported by IPMI is distributed between CPU and DRAM using RAPL counters. Then this total CPU and DRAM power usage is distributed among the running jobs based on the CPU time and memory usage of each job relative to the total CPU time and memory usage of the node. The final term represents the power usage by the network devices. CEEMS exporter does not export any network-related statistics at the moment and hence, the total power usage by networking is distributed equally among the running jobs at a given time. If power usage from other components like distributed storage, network infrastructure, cooling, \emph{etc.,} are available they can be easily integrated into the power estimation. This estimation formula can be customized based on underlying hardware using different Prometheus recording rules for different compute node groups. Example recording rules for different cases are provided in the \verb|etc/prometheus| folder of CEEMS GitHub repository~\cite{ceems:2024}.
	
	\subsection{Metrics visualization}
	
\begin{figure}[htbp]
		\subfloat[Aggregate usage metrics\label{fig:aggdash}]{\includegraphics[width=0.5\textwidth]{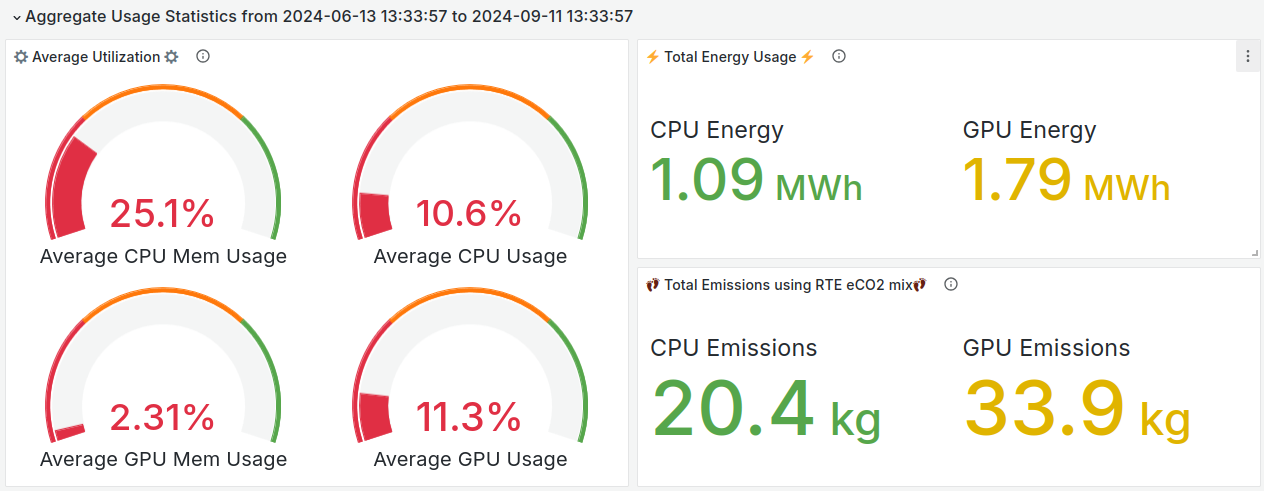}}\\
		\subfloat[List of SLURM jobs with aggregate metrics\label{fig:aggjobs}]{\includegraphics[width=0.5\textwidth]{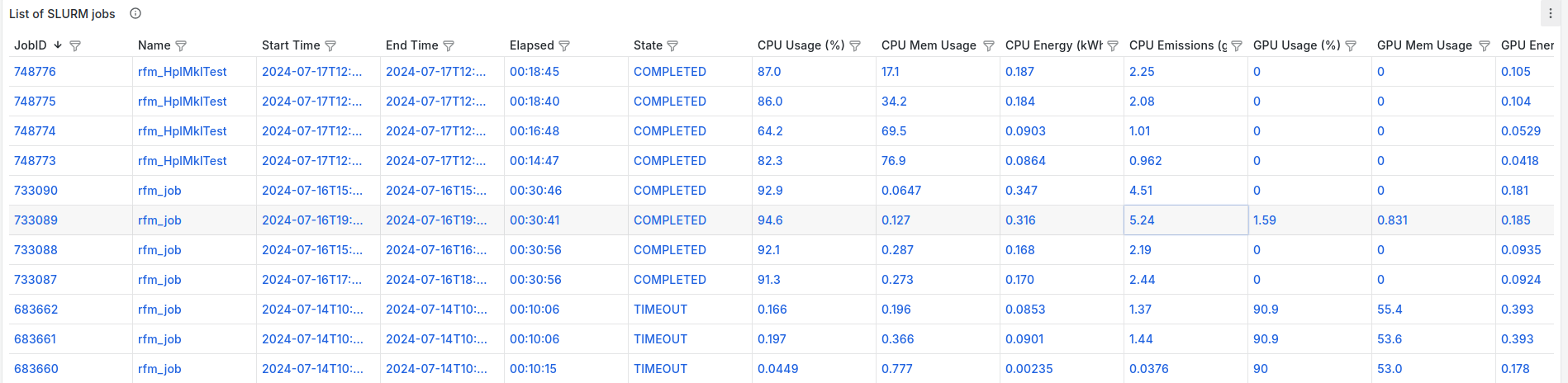}}\\
		\subfloat[Time series CPU metrics for a given job\label{fig:tsjob}]{\includegraphics[width=0.5\textwidth]{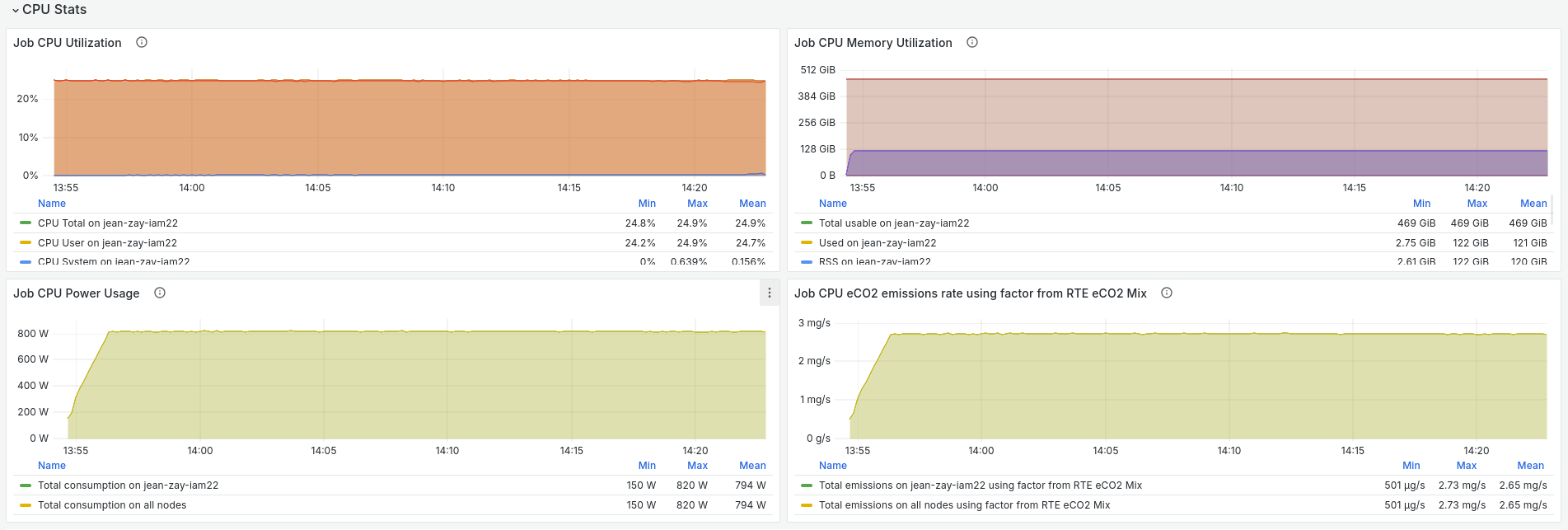}}
		\caption{Grafana dashboards of CEEMS.}
		\label{fig:dashes}
	\end{figure}

	CEEMS expose a variety of metrics, both time series and aggregated, which can help end users to keep track of either energy and environmental footprint and optimize the workloads. For example, Fig~\ref{fig:dashes} shows some of the dashboards built based on CEEMS on the Jean-Zay platform.
	Fig~\ref{fig:aggdash} shows the average CPU and GPU and their memory usage, total energy usage and resulting equivalent emissions of a user during the last 3 months. Fig~\ref{fig:aggjobs} lists SLURM jobs of a user along with aggregated metrics of each job. Finally, Fig~\ref{fig:tsjob} shows the time series CPU metrics of a given job.
	
	Cluster operators can have similar data available to them, albeit, for the entire cluster. This enables the operators to perform data analysis on the job metrics data to optimize the cluster usage, identify users and/or projects that are using the cluster resources inefficiently, \emph{etc.} 
	
	\section{Future work}
	\label{sec:4}
	
	CEEMS is still in the early development phase where initial feedback has been very encouraging. Some of the important features in the pipeline are adding network and IO stats to CEEMS exporter using extended Berkley Packet Filtering (eBPF) framework and adding performance metrics like FLOPS, caching, and memory IO bandwidth to name a few from Linux's \verb|perf| framework. Extending CEEMS to Openstack and Kubernetes is also a long-term objective.
	
	\section*{Acknowledgment}
	
	The author would like to thank the IDRIS' system team that authorized to install CEEMS on Jean-Zay and support team in providing insightful and useful feedback on CEEMS which helped to improve the overall CEEMS architecture.
	
	\bibliographystyle{IEEEtran}
	\bibliography{IEEEabrv,bibilography}
	
\end{document}